\documentclass[aps,prb,twocolumn,footinbib,showpacs,showkeys,superscriptaddress]{revtex4-2}
\usepackage{epsfig}
\usepackage{graphicx,graphics}
\usepackage{epstopdf}
\usepackage{natbib}
\usepackage{lipsum}
\usepackage{dcolumn}
\usepackage{amsmath,amssymb,amsfonts}
\usepackage{latexsym,verbatim}
\usepackage{bm}
\usepackage{color}
\usepackage{siunitx}
\usepackage[breaklinks=true,colorlinks,citecolor=blue,linkcolor=blue,urlcolor=blue]{hyperref}
\begin{document}
	
\title{Microscopic theory of photon-induced energy, momentum, and angular momentum transport in the nonequilibrium regime}
\author{Yong-Mei Zhang}
\affiliation{College of Science, Nanjing University of Aeronautics and Astronautics, Jiangsu 210016, People's Republic of China}
\author{Tao Zhu}
\email{zhutao@tiangong.edu.cn}
\email{phyzht@outlook.com}
\affiliation{Department of Physics, National University of Singapore, Singapore 117551, Republic of Singapore}
\affiliation{School of Electronic and Information Engineering, Tiangong University, Tianjin 300387, People's Republic of China}
\affiliation{Beijing National Laboratory for Condensed Matter Physics, Institute of Physics, Chinese Academy of Sciences, Beijing 100190, People's Republic of China}
\author{Zu-Quan Zhang}
\affiliation{Department of Physics, National University of Singapore, Singapore 117551, Republic of Singapore}
\affiliation{Department of Physics, Zhejiang Normal University, Jinhua 321004, Zhejiang, People's Republic of China}
\author{Jian-Sheng Wang}
\affiliation{Department of Physics, National University of Singapore, Singapore 117551, Republic of Singapore}
\date{\today}

\begin{abstract}
We set up a general microscopic theory for the transfer of energy, momentum, and angular momentum mediated by photons. Using the nonequilibrium Green's function method, we propose a unified Meir-Wingreen formalism for the energy emitted, force experienced, and torque experienced by the objects due to the fluctuating electromagnetic field. Our theory does not require the local thermal equilibrium that is the central assumption of the conventional theory of fluctuational electrodynamics (FE). The obtained formulas are valid for arbitrary objects as well as the environment without the requirement of reciprocity. To show the capability of our microscopic theory, we apply the general formulas to transport problems of graphene edges in both equilibrium and nonequilibrium situations. We show the local equilibrium energy radiation of graphene obeys the well-known $T^4$ law with a converged theoretical emissivity of 2.058$\%$. In the ballistic nonequilibrium situation driven by chemical potential biases, we observe nonzero results for force and torque from the graphene edges, which go beyond the predictive ability of the FE theory. Our method is general and efficient for large systems, which paves the way for studying more complex transport phenomena in the nonequilibrium regime.
\end{abstract}
\maketitle
\section{Introduction}
Due to fluctuations of the electromagnetic field around bodies, photons transfer energy, momentum, and angular momentum from one object to another or to the environment, which gives rise to abundant physical phenomena such as radiative heat transfer \cite{heat-volokitin,heat-biehs}, the Casimir force \cite{Casimir,lifshitz,casimir-woods}, and the associated torque \cite{torque-nature}. The transport problems of these conserved quantities have attracted tremendous interest for their myriad applications in advanced technologies. For example, the heat transfer in the near field can significantly exceed the black-body limit, which plays an important role in developing alternative techniques such as thermal management \cite{thermal-manage}, energy conversion \cite{energy-conversion}, data storage \cite{data-storage}, etc. Photon-carrying momentum generates force in both equilibrium and nonequilibrium situations. With certain geometries, the Casimir force arises, and novel phenomena such as levitation and the self-propelling state can be achieved \cite{kruger-prb2012,kruger-epl}. Angular momentum radiation plays a central role in quantum nanophotonics and topological electrodynamics \cite{AM1,AM2}. Recent studies demonstrated a separation of orbital angular momentum and spin angular momentum \cite{oam-sam1,oam-sam2}, which have been intensively applied to information and image processing \cite{AM3,AM4,AM5,AM6}.

Conventional theories for these transport phenomena are based on the fluctuational electrodynamics (FE) of Rytov \cite{rytov,pvh} which combines Maxwell's equations with the local fluctuation-dissipation theorem (FDT) \cite{FDT}. However, there are several situations in which the FE theory might fail. First, FE is a macroscopic theory with, usually, a phenomenological treatment of materials by a frequency-dependent local dielectric function. This is not sufficient at the subnanometer scale, for which detailed atomistic modeling of material properties is needed, especially for inhomogeneous materials or near edges \cite{kruger-review,zhu,microcasimir}. Second, applying the FDT requires a local thermal equilibrium for each object that is questionable in real experiments. Effort has been devoted to extending both the FDT and FE to nonequilibrium systems of multiple objects by combining the scattering theory with the conventional FE \cite{kruger-review}. However, the described system is still under a nonequilibrium stationary state in which each object has a distinct but still definite temperature. Furthermore, existing theoretical works focus on only one or two of these phenomena with the usual requirement of reciprocity \cite{heat,casimir1,casimir2,kruger-prl,both,zubin-prb}. To the best of our knowledge, a general unified microscopic theory for all three photon-induced transport phenomena, especially in the nonequilibrium regime, is still lacking.

In this paper, using the nonequilibrium Green's function (NEGF) method, we propose a general microscopic formalism for the photon-induced transfer of energy, momentum, and angular momentum in a unified fashion. We show that combined with the self-energy of the objects, the physical observables of three conserved quantities can be obtained from the corresponding quantum mechanical operators acting on the photon Green's function of the electromagnetic field. The obtained Meir-Wingreen-type formulas \cite{meir1,meir2} are valid for both objects and the environment without assumptions of local thermal equilibrium and reciprocity. In this regard, our theory allows for studying a different class of nonequilibrium transport phenomena beyond the applicability limit of existing FE approaches. In other words, the developed NEGF formalism in this paper can study transport problems for objects without a definite temperature, e.g., a ballistic nonequilibrium situation with driven currents.

To demonstrate the power of our microscopic theory, we study the edge effects of the transport phenomena of graphene nanoribbons in both equilibrium and nonreciprocal nonequilibrium situations. In particular, we calculate the energy emitted, force experienced, and torque experienced by graphene edges with possible electron transitions due to chemical potential biases. In thermal equilibrium, we show the edge effects of graphene have an approximate length scale of $t/k_BT$, where $t$ is the hopping parameter and $k_BT$ is the thermal energy at the temperature $T$. For the bulk two-dimensional system, the heat emission of graphene obeys the $T^4$ law with a converged emissivity of $2.058\%$, which is in good agreement with the value implied by the Dirac model \cite{Falkovsky}. Moreover, we demonstrate nonzero momentum and angular momentum radiations in a ballistic nonequilibrium situation which cannot be treated by the conventional FE. The discovered nonvanishing force and torque at the edge are unique from nonequilibrium steady states.

\section{Theory}
To tackle the fluctuating electromagnetic field around bodies, we explore the NEGF method \cite{negf1,negf2,review} as our basic tool. The fundamental quantity of interest is the photon Green's function defined by the vector potential $A_\mu$ as
\begin{equation}
	\label{eq-greenD}
	D_{\mu \nu}({\bf r}, \tau ; {\bf r}',\tau') = \frac{1}{i \hbar} \bigl\langle T_\tau A_{\mu}({\bf r}, \tau) A_{\nu}({\bf r}', \tau') \bigr\rangle. 
\end{equation}
Here $\tau$ and $\tau'$ are Keldysh contour times, ${\bf r}$ and ${\bf r}'$ are the positions,  $T_\tau$ is the contour-order operator, and $\mu$ and $\nu$ take $x$, $y$, and $z$ directions. The average $\langle...\rangle$ shown in Eq.~(\ref{eq-greenD}) is a nonequilibrium average by a certain unknown density matrix whose effect can be reflected by the properties of the baths. From the contour Green's function, we can determine the lesser ($<$), greater ($>$), retarded ($r$), and advanced ($a$) Green's functions in the usual way as defined in Appendix A. For convenience, we adopt the $\phi = 0$ gauge \cite{Heisenberg_pauli} with electric field strength ${\bf E} = - \partial {\bf A}/\partial t$ and magnetic induction by ${\bf B} = \mbox{\boldmath{$\nabla$}} \times {\bf A}$. A perturbation theory with the $-{\bf j} \cdot {\bf A}$ interaction, where ${\bf j}$ is the electric current density, leads to a Dyson equation $D = v + v \Pi D$, where $v^{-1} = \epsilon_0 (\omega^2 - c^2 \nabla \times \nabla \times \cdot)$ is a differential operator acting on $D$ in the frequency domain, $c$ is the speed of light, and the self-energy $\Pi$ is the lowest-order current-current correlation function in the random phase approximation.

%problem setup

\begin{figure}
	\includegraphics[width=8.6 cm]{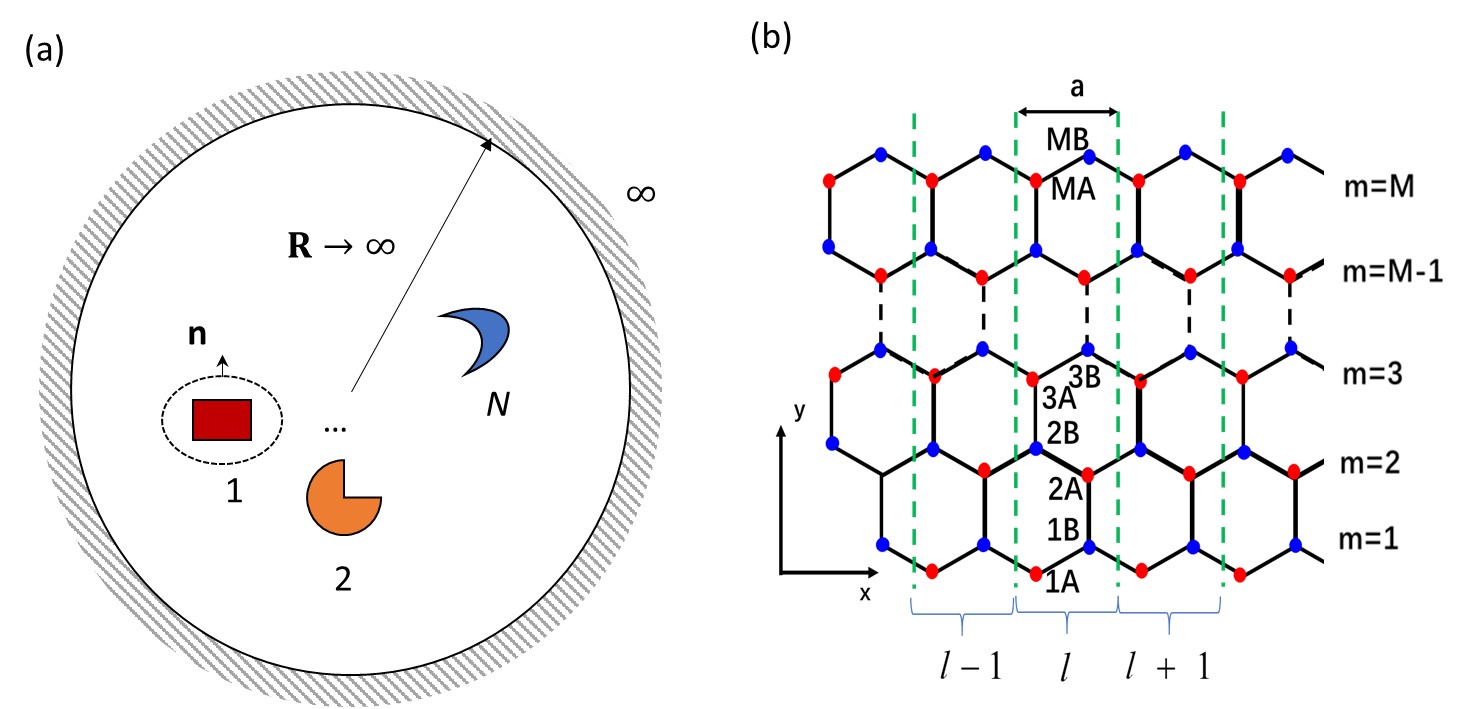}
	\caption{(a) Schematic setup of $N$ objects in vacuum with arbitrary geometry and the environment served as bath at infinity. (b) Graphene nanoribbon with zigzag edges. The $x$ direction is periodic, and unit cells are represented by dashed lines with index $l$. Each site along the $y$ direction is labeled by index $m$ with two sublattice sites $A$ and $B$.}
	\label{fig1}
\end{figure}

Now, we consider $N$ physical objects with arbitrary geometry in vacuum. The environment is represented by a sphere of ``bath at infinity'' with radius ${\bf R} \to \infty$, as depicted in Fig.~\ref{fig1}(a). The major physical observables of energy, momentum, and angular momentum transport are the heat emitted, radiation resultant force, and torque applied to each of the $N$ objects, respectively. We calculate these observables by surface integration of the corresponding fluxes with an outward norm. Using the divergence theorem, the surface integrals can be transformed into volume integrals over the object. Then, we have the net energy emission $I_\alpha$, the total force experienced ${\bf F}_\alpha$, and the total torque experienced ${\bf N}_\alpha$ for object $\alpha$:
\begin{equation}
	\label{em}
	\left(\!\!\begin{array}{c}
		I_\alpha \\ {\bf F}_\alpha \\ {\bf N}_\alpha
	\end{array}\!\!
	\right) = \oint_{\bf \Sigma} \left(\!\!\begin{array}{c}
		{\bf S} \\ {\bf T} \\ {\bf r}\times{\bf T}
	\end{array}\!\! \right)\cdot d{\bf \Sigma}
    = \int_V \left(\!\!\begin{array}{c}
    	-{\bf E}\cdot{\bf j} \\ {\bf f} \\ {\bf r}\times{\bf f}
    \end{array} \!\! \right) dV,    
\end{equation}
where  $d{\bf \Sigma}$ is the surface element with an outward norm, $dV$ is the volume element, ${\bf S} = ({\bf E} \times {\bf B})/\mu_0$ is the Poynting vector, ${\bf T} = \epsilon_0 {\bf E} {\bf E} + \frac{1}{\mu_0} {\bf B} {\bf B} - u {\bf U}$ is the Maxwell stress tensor with $u = \frac{1}{2} (\epsilon_0 E^2 + B^2/\mu_0)$, ${\bf U}$ is the identity, and ${\bf f} = \rho {\bf E} + {\bf j} \times {\bf B}$ is the Lorentz force density. Since charge and current densities are related by the continuity equation $\partial\rho/\partial t = -\nabla\cdot{\bf j}$, at steady state, we can perform integration by parts in time as well as in space [namely, $\langle a \dot{b} \rangle = - \langle \dot{a} b \rangle$, $\int dV a\, \partial_\mu b = -\int dV (\partial_\mu a)\, b$], transforming the force density as ${\bf f} = \sum_{\nu} j_\nu \mbox{\boldmath{$\nabla$}} A_\nu$.

In performing the quantum-mechanical steady-state average, we use a symmetric order of the operators, $\langle A B  + B A \rangle/2 = i\hbar\int_0^\infty  \frac{d\omega}{2\pi} G_{AB}^K(\omega)$, where the Keldysh Green's function is $G^K = G^<+G^>$, with $G_{AB}^<(t-t') = \langle B(t')A(t) \rangle/(i\hbar)$ and $G_{AB}^>(t-t') = \langle A(t)B(t') \rangle/(i\hbar)$. To conveniently manipulate the expression, we introduce an intermediate quantity
\begin{equation}
	F_{\mu\nu}({\bf r},\tau; {\bf r}', \tau') = \frac{1}{i\hbar} \bigl\langle T_\tau A_\mu({\bf r},\tau) j_\nu({\bf r}',\tau')\bigr\rangle.
\end{equation}
Then, the current density $\bf j$ connects back to the vector field $\bf A$ by evoking the linear response, ${\bf j} = - \Pi_\alpha {\bf A}$ on the contour. Since the $\langle AA\rangle$ correlation gives the Green's function $D$, after applying the Langreth rule \cite{langreth} to $D\Pi_\alpha$ defined on the contour, we can express the three observables in terms of $F$, by taking time or spatial derivatives, for which more details can be found in Appendix B.

Finally, we obtain our central results, which are the following Meir-Wingreen type formulas for ${I_\alpha}$, ${\bf F}_\alpha$, and ${\bf N}_\alpha$:
\begin{equation}
	\label{eq-main-result}
	\left(\!\!\begin{array}{c}
		I_\alpha \\ {\bf F}_\alpha \\ {\bf N}_\alpha
	\end{array}\!\!
	\right) = \int_0^\infty \!\! \frac{d\omega}{2\pi} 
	{\rm Re}{\rm Tr}\left[ \left(\!\!\begin{array}{c}
		-\hbar\omega \\ \hat{\bf p} \\ \hat{\bf L}
	\end{array}\!\! \right) 
	\bigl( D^r \Pi^K_\alpha + D^K \Pi^a_\alpha\bigr) \right].
\end{equation}
In the above, $\hbar\omega$ is obtained from the energy operator $i\hbar\partial/\partial t$ acting on the Fourier transform of $D$; $\hat{\bf p} = -i \hbar \mbox{\boldmath{$\nabla$}}$ is the momentum operator acting on the first spatial argument of $D$. $\hat{\bf L} = {\bf r} \times \hat{\bf p} + \hat{{\bf S}}$ is the angular momentum operator with the spin operator $S^\mu_{\nu\gamma} = (-i \hbar) \epsilon_{\mu\nu\gamma}$ acting on the directional index space of $D$. $\epsilon_{\mu\nu\gamma}$ is the Levi-Civita symbol. The trace is over the space ${\bf r}$ as a volume integral and summation over the index $\mu$. ${\rm Re}$ stands for the real part. The lesser (greater) Green's function is related to the retarded and advanced ones with the Keldysh equation $D^{<(>)} = D^r \Pi^{<(>)} D^a$, where $\Pi ^{<(>)}$ is the total self-energy summed over $\alpha$, i.e., $\Pi^{<(>)} = \sum_{\alpha}\Pi_\alpha^{<(>)}$. In a tight-binding model, if we use an electron system with the electron-photon coupling matrix $M_{jk}^{l\mu}$ for the unit cell $l$, with $j$ and $k$ being electron sites, the self-energy from each object, serving also as bath for photons, is given by
\begin{equation}
	\Pi^{l\mu,l'\nu}_\alpha(\tau,\tau') = -i\hbar\, {\rm Tr}_e\big[ M^{l\mu} G(\tau, \tau') M^{l'\nu}G(\tau',\tau)\bigr],
\end{equation}
where $G$ is the electron Green's function of object $\alpha$ and the trace is over the electron sites. We note that the retarded self-energy is related to the dielectric function by $\Pi^r = - \epsilon_0 \omega^2 (\epsilon - 1)$, which in turn can be related to the electric conductivity of materials \cite{zhu2}. 

Another advancement of the Meir-Wingreen formulas Eq.~(\ref{eq-main-result}) is that they also works for the environment ($\alpha = \infty$) where the trace operation is interpreted as an integration over the sphere and the sum over the direction index. The self-energy for the bath at infinity can be worked out by conservation laws of the three kinds of transport quantities for $\alpha = 1, 2, \cdots, N$ and $\infty$ as a whole.  This requirement demands $\Pi_{\infty}^r = - v^{-1}$. Alternatively, a dust model integrating over the space $| {\bf r} | > R$ can be built by considering a one-dimensional chain model with a finite central region and considering the effect of non-reflecting boundary conditions \cite{peng}. Finally, one matches the surface integral results of the Poynting vector and Maxwell stress tensor. They all lead to the expression
\begin{equation}
	\label{eq-bath-at-inf}
	\Pi^r_{\infty} = - i \epsilon_0 c\, \omega \left( {\bf U} - \hat{\bf R} \hat{\bf R} \right),
\end{equation} 
where the retarded self-energy is expressed in dyadic notation and $\hat{\bf R} = {\bf R}/R$ is the unit vector pointing from the coordinate origin to a point on the sphere.

From the above discussion, it appears that we have two self-energy expressions for the bath at infinity. One is $-v^{-1}$, and the other is given by Eq.~(\ref{eq-bath-at-inf}). The former is a differential operator which must act on $D$ in order to see its effect, and the latter is defined on the sphere $|{\bf r}| = R$, which is friendlier for actual computation. Here, we demonstrate their consistency with a dust model. We assume for $|{\bf r}| \ge R$ that the solution for the retarded Green's function is the free one, 
\begin{eqnarray}
	D^r_0 &=& - \frac{e^{i\frac{\omega}{c}R}}{4\pi \epsilon_0 c^2 R} \\
\nonumber&\times&\left[({\bf U} - \hat{\bf R}\hat{\bf R}) + \left( - \frac{1}{i\frac{\omega }{c}R} +\frac{1}{(i \frac{\omega}{c} R)^2} \right)({\bf U} - 3\hat{\bf R}\hat{\bf R}) \right].
\end{eqnarray}
A ``dust" model is obtained by the replacement $\omega \to \omega + i \eta$ in the above solution to describe the damping for
$|{\bf r}| > R$. In evaluating the transport quantities, we need to evaluate the trace of the form
${\rm Tr}\, \hat {O} \bigl[  D^r \Pi^< D^a \Pi^a_\infty \bigr]$ in which $\hat{O}$ is the extra operator acting on $D^r$, and the trace involving the bath at infinity is the volume integral of all space outside the sphere. Then we can perform a solid angle integration and $\int_R^\infty dr\, r^2 \cdots$. Since when $\eta = 0$, $v^{-1} D^r = 0$, the effect of the dust is 
\begin{equation}
	v^{-1} D^r = \epsilon_0 \left[ \omega^2 - (\omega + i\eta)^2\right] D^r \approx \epsilon_0 ( - 2 i \eta \omega) D^r.
\end{equation}
The advanced Green's function $D^a$ is obtained by taking the Hermitian conjugate of $D^r$. At an asymptotically large distance, we can ignore the second
high-order term in $1/R$ and it is sufficient to keep the first term. Then the decay factor is $D^r D^a \propto e^{-2 \eta r/c}$.
After integrating $r$ from $R$ to infinity, we find a finite result when $\eta \to 0^+$, 
\begin{equation}
	{\rm Tr}\, \hat {O} \bigl[  D^r \Pi^< D^a \Pi^a_\infty \bigr] \approx
	\int d\Omega R^2 \hat {O} \bigl[  D^r \Pi^< D^a ( i \epsilon_0 c \omega) \bigr].
\end{equation}
In the process, the operator $\hat{O}$ should not mess up the argument. Both $D^r$ and $D^a$ have a transverse projector ${\bf U} - \hat{\bf R}\hat{\bf R}$, so we also attach this projector to the numerical factor $i \epsilon_0 c \omega$.
This gives the surface sphere version of the self-energy for the bath at infinity in Eq.~(\ref{eq-bath-at-inf}), which is local in the solid angle. 

As a check of the correctness of the self-energy, we consider a system that consists solely of the bath at infinity and no objects at all. We then evaluate the energy density $u = \frac{1}{2} ( \epsilon_0 E^2 + \frac{1}{\mu_0} B^2)$ at the origin,
due to the bath at infinity with temperature $T$. The thermal average can be expressed in terms of the Green's function as
\begin{equation}
	\langle u \rangle = \int_0^\infty \!\! \frac{d\omega}{2\pi} i\hbar  {\rm Tr}_{\mu}\left[
	\epsilon_0 \omega^2 D^< -  
	\frac{1}{\mu_0} \mbox{\boldmath{$\nabla$}}_{{\bf r}} \!\times\! D^< \!\times\! \mbox{\boldmath{$\nabla$}}_{{\bf r}'}  
	\right],
\end{equation}
where the trace is in the direction index, the first gradient operator is over the first argument, and the second one is over the second argument. After taking the derivatives, the Green's function is evaluated at ${\bf r} = {\bf r}' = \bf 0$. By applying the Keldysh equation, $D^< = D^r \Pi^<_\infty D^a$, and the fluctuation-dissipation theorem, $\Pi^<_\infty = N(\omega) \bigl( \Pi^r_\infty - \Pi^a_\infty\bigr)$, we can perform the trace at the sphere while expressing $D^r_0$ to the leading order in $1/R$. Then we obtain
\begin{equation}
	\langle u \rangle = \int_0^\infty d\omega\,  \frac{\omega^2}{\pi^2 c^3} \,  \hbar \omega\, N(\omega),
\end{equation}
which is the correct expression for the blackbody radiation. Interestingly, an explicit calculation from Eq.~(\ref{eq-main-result}) shows that, for the far-field angular momentum emission, exactly half is from the orbital contribution ${\bf r}\times{\bf p}$, and the other half is from the spin part, in which the final expression agrees with the surface integral result \cite{zuquan-prb}. 

\section{example of graphene nanoribbon}
Now, we apply the general theory to the system of a graphene nanoribbon, as shown in Fig.~\ref{fig1}(b). In particular, we study the nonequilibrium edge effects in a nonreciprocal ballistic transport situation in which the conventional FE theory fails. For the convenience of saving computational cost, we perform the following approximations for a concrete calculation. First, we make a multipole expansion,  
$D^r({\bf R}, {\bf r}) = D^r({\bf R}, {\bf 0}) + {\bf r} \cdot \partial D^r({\bf R}, {\bf r}')/\partial {\bf r}'|_{{\bf r}'={\bf 0}}+ \cdots$, and keep the dominant lowest non-vanishing order (note that the monopole term for force is identically zero). Second, the graphene ribbon is lattice periodic in the $x$ direction, so the self-energy can be calculated in the eigenmode representation, which is more efficient than that of frequency integration \cite{zuquan-prb}. After performing the integration of the solid angle and frequency as shown in Appendix C, we obtain the formula of emitted energy as
\begin{eqnarray}
\label{energy}
\nonumber I &=& \dfrac{4\alpha}{3\hbar c^2}\sum_{\mu,nn'}(\varepsilon_n-\varepsilon_{n'})^2\Theta(\varepsilon_n-\varepsilon_{n'})\lvert\langle n|V^\mu|n'\rangle\rvert^2\\
 &\times&f_n(1-f_{n'}),
\end{eqnarray}
where $\alpha \approx 1/137$ is the fine structure constant and $\varepsilon_{n}$ is the energy of state $n$ with the Fermi distribution function $f_{n}=1/(e^{\beta_{L(R)}(\varepsilon_{n}-\mu_{L(R)})}+1)$, where $\beta_{L(R)}=1/k_BT_{L(R)}$ and $\mu_{L(R)}$ is the chemical potential applied to the left (right) lead. Taking the left or right chemical potential in the Fermi function is determined by the sign of the group velocity, which is calculated by $\langle n|V^x|n\rangle$ \cite{Datta}, where $V^\mu$ is the component of the velocity matrix in the $\mu$ direction (see details in Appendix D). This treatment realizes a nonequilibrium situation in a ballistic system. $\Theta(x)$ is the step function, which is 1 for $x>0$ and 0 otherwise. Not surprisingly, this formula agrees with results obtained from Fermi's golden rule and the Boltzmann transport theory \cite{fgr}. Similarly, we can derive the formula for torque as
\begin{eqnarray}
N_z \nonumber&=&\dfrac{4\alpha}{3c^2}\sum_{nn'}(\varepsilon_n-\varepsilon_{n'})\Theta(\varepsilon_n-\varepsilon_{n'}){\rm Im}\Bigg\{f_n(1-f_{n'})\\
&\times&\Big[\langle n|V^x|n'\rangle\langle n'|V^y|n\rangle-\langle n|V^y|n'\rangle\langle n'|V^x|n\rangle \Big]\Bigg\},
\end{eqnarray}
which is also consistent with the real-space formula reported in Ref. \cite{zuquan-prb}.

The situation for force is a bit more complex in that a concrete formula has not yet been given. Here we introduce the notation $eU^\mu_\gamma\equiv\sum_l{ M}^{l\mu}r^l_\gamma$. The superscript index is associated with ${ M}^{\mu l}$ or the velocity component, and the subscript index is associated with the direction of coordinate $r_\gamma$. With some derivations for which the details can be found in Appendix C, we obtain the force formula
\begin{eqnarray}
\nonumber F^\mu &=&\dfrac{4\alpha}{30\hbar^2 c^4}\sum_{nn'}\left(\varepsilon_{n}-\varepsilon_{n'}\right)^3\Theta(\varepsilon_{n}-\varepsilon_{n'})f_n(1-f_{n'})\\
 \nonumber&\times&{\rm Tr}\left[4\sum_\nu(\rho_nU^\nu_\mu\rho_{n'}V^\nu-\rho_nV^\nu\rho_{n'}U^\nu_\mu)\right.\\
 \nonumber&&\qquad-\sum_\nu(\rho_nU^\mu_\nu\rho_{n'}V^\nu-\rho_nV^\mu\rho_{n'}U^\nu_\nu)\\
&&\qquad-\left.\sum_\nu(\rho_nU^\nu_\nu\rho_{n'}V^\mu-\rho_nV^\nu\rho_{n'}U^\mu_\nu)\right],
\end{eqnarray}
where $\rho_n = |n\rangle\langle n|$ is the density matrix of state $n$.

\begin{figure}
	\centering
	\includegraphics[width=8.6cm]{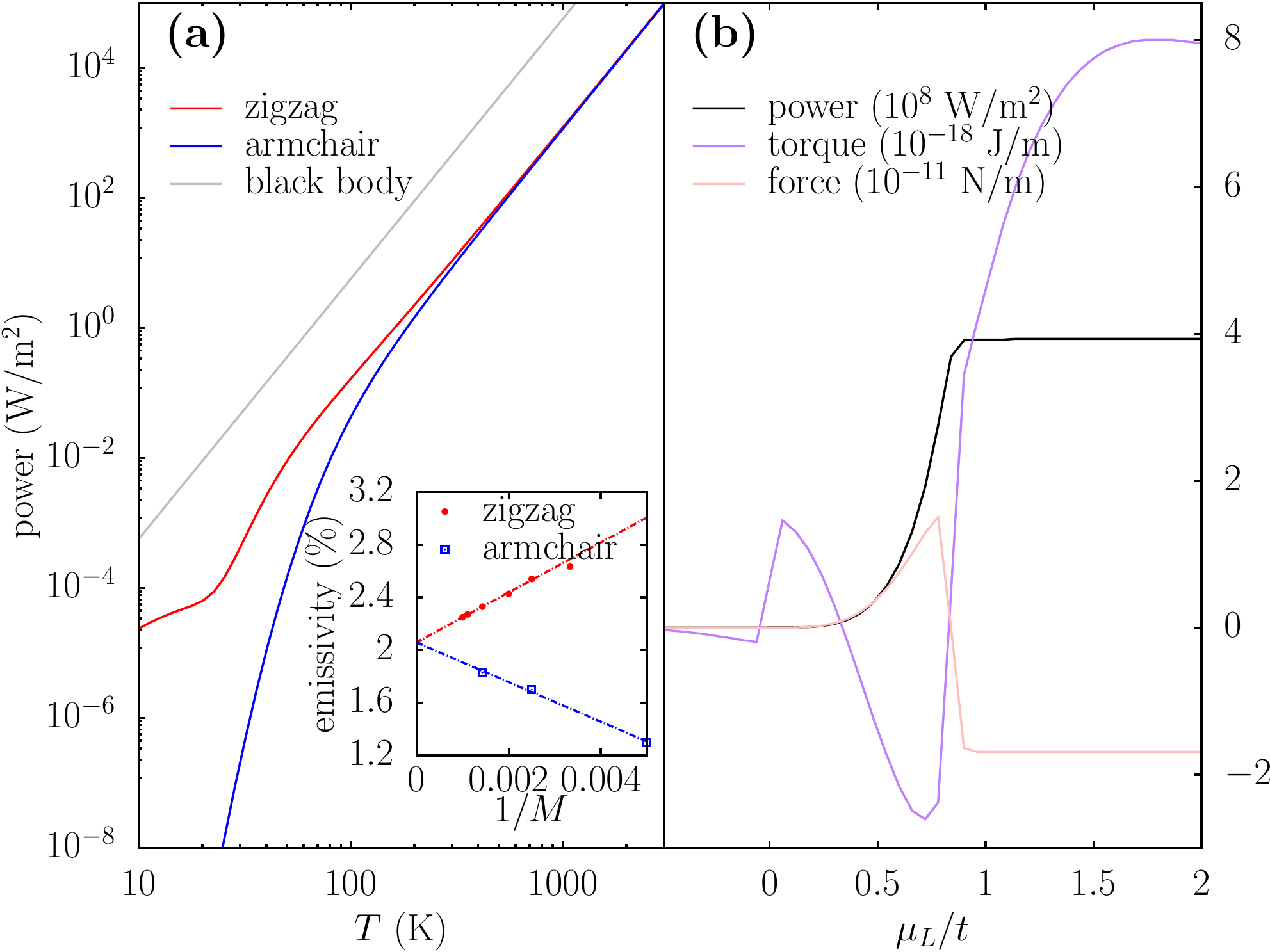}
	\caption{(a) Temperature dependence of radiation power from zigzag and armchair graphene nanoribbons. Inset: the calculated emissivity as a function of the inverse of the ribbon width at the temperature of 300~K. (b) Photon-induced power emitted, torque experienced, and force experienced by graphene nanoribbons as functions of the left lead chemical potential with 401 $k$ points at a temperature of 300~K. The right lead chemical potential is fixed at $\mu_R=-0.84t$.}
	\label{fig2}
\end{figure}

%\onecolumngrid
\begin{figure*}[t]
	\includegraphics[width=0.9 \textwidth]{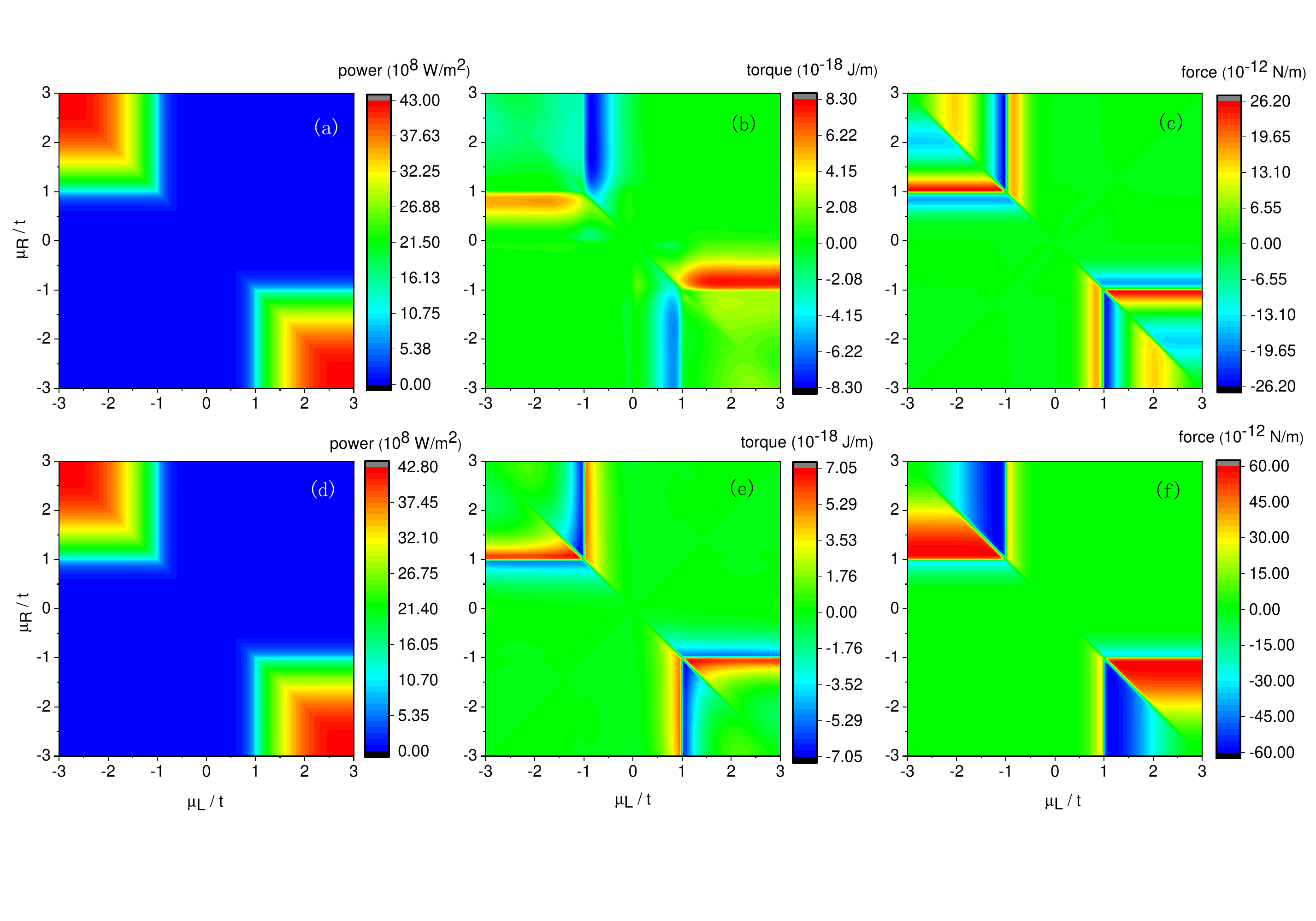}
	\caption{Intensity of (a) power, (b) torque, and (c) force for zigzag graphene nanoribbon edges with different chemical potentials for the left and right leads. (d)-(f) The same quantities as in (a)-(c), respectively, but for armchair graphene nanoribbon edges.}
	\label{fig3}
\end{figure*}
%\twocolumngrid

Now, we first study the edge effects of the energy emission of graphene nanoribbons in thermal equilibrium. We show in Fig.~\ref{fig2}(a) the thermal radiations from graphene nanoribbons with stripe width $M=700$. As shown, the emission powers from both zigzag and armchair ribbons converge at high temperatures with the $T^4$ law of gray bodies. The difference between the zigzag and armchair ribbons at low temperature is because the former is metallic while the latter has a very small bandgap. With the increase in temperature, thermal excitation gradually overcomes the small energy gap, and their difference vanishes. On the other hand, as the edge and bulk contribute differently to the energy emission, we see deviations of the $T^4$ law for both ribbons at low temperature due to the finite stripe used. This is further supported by the fact that, for $t = 2.8$~eV, the width $M=700$ is not much larger than $t/k_BT\approx100$ at room temperature. In the inset in Fig. 2(a),  we check the convergence of the edge effects of the zigzag and armchair ribbons. We find the emissivities of both zigzag and armchair ribbons have an inverse linear relation to the width. For infinitely large width (i.e., $1/M\to0$), the emissivity converges to $2.058\%$, matching almost exactly to the emissivity of $2.056\%$ obtained analytically from integration over the whole frequency range of light using the Dirac model \cite{Falkovsky}. This value is independent of temperature and also agrees with the experimental reports \cite{marcus} that the thermodynamic emissivity of graphene is smaller than the $2.3\%$ optical emissivity obtained from a circle average of the visible light radiation \cite{nair2008fine}.

According to the FE theory, torque and force are zero for isolated bulk materials due to the reciprocity of the Hamiltonian \cite{kruger-bulk}. Thus, we further study the transport phenomena of the graphene edge under a ballistic nonequilibrium situation which the FE theory cannot handle. We show in Fig.~\ref{fig2} (b) the calculated power emitted, torque experienced, and force experienced by graphene nanoribbons under different chemical potential biases. Specifically, the chemical potential biases are applied in the $x$ direction, and we compute the torque in the $z$ direction and force in the $x$ direction as all other directions generate null results. As shown, power, torque, and force are significantly changed with $\mu_{L}$ altered from 0 to $t$, while all remain unchanged for $\mu_{L}$ outside of this range. This is attributed to the concentration of the density of states in this energy window. Interestingly, the energy emission increases significantly with the increase of $\mu_{L}$ from 0 to $t$, but the torque and force have more complex features. At certain ranges, the torque and force decrease with increasing $\mu_{L}$ due to the cancellation of momentum and angular momentum radiations from transitions between the energy state $0\to t$ and $0 \to-t$.

The effect of the chemical potential can be seen more clearly in the density plot. In Fig.~\ref{fig3}, we show power, torque, and force density of both zigzag and armchair graphene ribbons with chemical potentials $\mu_{L}$ and $\mu_{R}$. Generally, equal chemical potentials in the left $\mu_{L}$ and right $\mu_{R}$ lead do not generate torque or force. As shown in Figs.~\ref{fig3}(a) and (d), the energy radiations from zigzag and armchair ribbons are almost identical. However, the force and torque show very different patterns for zigzag and armchair edges. In Figs.~\ref{fig3}(b) and (e), stripe patterns can be observed in the diagonal corner. These stripes represent large angular momentum radiation under different chemical potential biases, corresponding to emission due to electronic transitions between the Van Hove singularities at $\pm t$ and zero-energy edge states. The forces shown in Figs.~\ref{fig3}(c) and (f) have similar features but more complicated patterns. This character is ascribed to the existence of zero-energy edge states of a zigzag graphene ribbon, while for an armchair ribbon, resonant transitions occur between the Van Hove singularities. Thus, the nonvanishing torques and forces emerging at nonequilibrium are edge effects in which different results for zigzag and armchair ribbons are shown.

\section{conclusion}
In summary, we have proposed a general theory for transport problems of conserved quantities mediated by photons. Using the NEGF method, we derived unified Meir-Wingreen type formulas for the energy emitted, force experienced, and torque experienced of objects (or environment) due to the fluctuating electromagnetic field. The theory is valid in both equilibrium and nonequilibrium regimes without the requirement of local thermal equilibrium and reciprocity of materials. We apply the general theory to the near-edge transport problem of graphene nanoribbons. In thermal equilibrium, the energy emission of graphene follows the $T^4$ law with an emissivity of 2.058\% for both infinitely wide zigzag and armchair ribbons. Then, we set up a nonequilibrium state in the ballistic regime by putting different chemical potentials based on group velocity in the $x$ direction. Our results show that nonzero momentum and angular momentum radiation can be generated from the edge in such nonequilibrium situations. There are prominent changes in all three conserved quantities in the bias window $0<\mu_{L}<t$, which goes beyond the predictability of the conventional FE theory. Regardless of the theoretical complexity, the proposed Meir-Wingreen formulas are general and can be applied to multiple objects to study more complex phenomena such as nonequilibrium Casimir effects in future work.

\section*{acknowledgment}
This work is supported by the Ministry of Education, Singapore, under its MOE tier 2, Grant No. R-144-000-411-112. 

\section*{Appendix A: Green's functions}
\setcounter{equation}{0}
\makeatletter
\renewcommand{\theequation}{A\arabic{equation}}

It is simpler to use a compact notation so that the Green's function defined by Eq. (1) in the main text is a matrix indexed by the space location ${\bf r}$ and index $\mu$, denoted $D(\tau, \tau')$.  The contour time is the pair $\tau = (t,\sigma)$ of real time and the branch index.  Due to the $+$ (forward) and $-$ (backward) branches the contour Green's function gives four Green's functions in real time: $D^{++} = D^t$ is time ordered, $D^{--}  = D^{\bar{t}}$ is anti-time ordered, $D^{+-} = D^<$ is lesser, and $D^{-+} = D^>$ is greater.   The four are not linearly independent and are constrained by  $D^t + D^{\bar{t}} = D^> + D^< = D^K$. The retarded Green's function is $D^r = D^t - D^< = \Theta(D^> - D^<)$, and the advanced Green's function is $D^a =D^< - D^{\bar{t}} = - (1-\Theta)(D^> - D^<)$, such that $D^> - D^< = D^r - D^a$. Let $1 \equiv ({\bf r}, \mu, t)$ and $2 \equiv ({\bf r}', \nu, t')$; we have the symmetry in the time domain as $D^>(1,2) = D^<(2,1)$ and $D^r(1,2) = D^a(2,1)$. The Fourier transform into frequency is defined by
\begin{equation}
	D(\omega) = \int_{-\infty}^{+\infty} dt\, D(t-t') e^{i \omega (t-t')}.
\end{equation}
In the frequency domain, we have the Hermitian conjugate  $[D^r(\omega)]^\dagger = D^a(\omega)$, $[D^<(\omega)]^\dagger = - D^<(\omega)$.
These general relations are also shared by the self-energy $\Pi$, since $\Pi$ is essentially the current-current Green's function. We define reciprocal as being $\Pi^{\rm T} = \Pi$, where the transpose is in the combined (${\bf r}, \mu$) space.

The contour-ordered Dyson equation, $D = v + v \Pi D$, implies the Keldysh equation, $D^< = D^r \Pi^< D^a$, which is valid in general. In global thermal equilibrium, we also have the fluctuation-dissipation theorem, $D^< = N(\omega) (D^r-D^a)$, where $N(\omega) = 1/(e^{\beta \hbar \omega} - 1)$ is the Bose function. Consistency between the two equations and with the retarded Dyson equation, when the self-energies are additive among the $N+1$ objects, requires  the self-energy for the bath at infinity to be  $\Pi^r_\infty =- (v^r)^{-1}$  (actually, the argument determines only the difference $\Pi^r - \Pi^a$).
This is because the Keldysh equation, defined in a compact domain (central region), together with the fluctuation-dissipation
relation, implies
\begin{equation}
	(D^a)^{-1} - (D^r)^{-1} = \sum_{\alpha=1}^{N+1}(\Pi^r_\alpha - \Pi^a_\alpha),
\end{equation} 
where we denote the ``object at infinity'' as $N+1$,  while the Dyson equation is
\begin{equation}
	(D^r)^{-1} = (v^r)^{-1} - \sum_{\alpha=1}^N \Pi^r_\alpha.
\end{equation}
The retarded Dyson equation should be viewed as a differential equation defined on the whole space, and the ``bath at infinity'' is only a no-scattering boundary condition at infinity, thus there is no explicit bath self-energy in the Dyson equation. 
Taking the Hermitian conjugate, subtracting the retarded version, and making a comparison, we find 
$\Pi^r_\infty - \Pi^a_\infty = (v^a)^{-1} - (v^r)^{-1}$.  We can fix it as an equality, $\Pi^r_\infty = -(v^r)^{-1}$, by conservation laws.
\section*{appendix B: Derivation of the Meir-Wingreen formulas}
\setcounter{equation}{0}
\makeatletter
\renewcommand{\theequation}{B\arabic{equation}}
We could follow the steps of Kr\"uger \textit{et al.} \cite{kruger-prb2012}, using ${\bf j} = - v^{-1} {\bf A}$ to map back to the field ${\bf A}$.  Here we use the method outlined in the main text.  The advantage of this route is that it is easy to separate out the total contribution from a focused object $\alpha$.  Using our definition of the Green's function $F=\langle Aj_\alpha\rangle/(i\hbar)$ and the expressions for the Joule heating, force, and torque formulas, we obtain
\begin{eqnarray}
	I_\alpha &=& \int_0^\infty \frac{d\omega}{2\pi} \hbar \omega \,{\rm Tr} \left[ F^K(\omega) \right], \\
	{\bf F}_\alpha &=&  \int_0^\infty \frac{d\omega}{2\pi} i \hbar \,{\rm Tr} \left[ \nabla_{{\bf r}} F^K(\omega) \right], \\
	{\bf N}_{\alpha} &=&  \int_0^\infty \frac{d\omega}{2\pi}   {\rm Tr} \Big[ i \hbar\, {\bf r} \times \nabla_{{\bf r}} F^K(\omega) -\hat{{\bf S}} F^{K}(\omega) \Big].
\end{eqnarray}
Here the differentiation with respect to time is $-i\omega$ in the frequency domain, and the differential in space is associated with $\bf A$, which is the first argument of $F({\bf r}, {\bf r}')$. The trace means integration over the whole volume and sum over the direction index. $\hat{\bf S}$ is the spin-1 operator, as explained in the main text. This extra spin term is caused by $- \int dV {\bf r} \times \partial_\nu (j_\nu {\bf  A}) = \int dV {\bf j} \times {\bf A}$, which is a total divergence for force but not for torque.

The next step is to evaluate the Green's function $F$. One can use the Feynman-diagrammatic method, but here we use a fast argument. The local response of the current due to total field is given by ${\bf j} = - \Pi_\alpha {\bf A}$, where both the current and field are interpreted as quantum operators, while the contour-ordered Green's function $\Pi$ is just a number.  The multiplication is a convolution in ${\bf r}$, $\tau$, and matrix multiplication in index $\mu$.  Putting this result into $F$, and using the fact that the contour-ordered Green's function for $D$ and $\Pi$ is symmetric with respect to the arguments, we obtain  $F = - D\Pi_\alpha$. The lesser (greater) component is obtained from Langreth's rule, $-F^{<(>)} = D^r \Pi^{<(>)}_\alpha + D^{<(>)} \Pi^a_\alpha$, which is then our main result. If we assume reciprocity for $\Pi$, we can recover the results for energy and force from Kr\"uger \textit{et al} \cite{kruger-prb2012}.

As we have defined the three conserved quantities as surface integrals, it is obvious that if we sum over the objects from $\alpha = 1, 2, \cdots, N$, and $N+1$, we should get zero. The conservation of energy is obtained from the identity ${\rm Tr}(D^> \Pi^< \!-\! D^< \Pi^>)=0$, where $\Pi^{>,<}$ is the total. This identity cannot be used to prove the conservation of total force and torque due to the extra operator in front of $D$, but the sum is indeed zero if  $\Pi_\infty^a = -(v^a)^{-1}$. This is a consequence of the validity of the Dyson equation. After the summation, the last common factor in Eq.~(4) after the various operators is $D^r \Pi^K + D^K \Pi^a = D^r \Pi^K ( I + D^a \Pi^a)$, and we have used the Keldysh equation. The extra multiplicative factor $I + D^a \Pi^a$ when equated to zero is nothing but the Dyson equation of the advanced version, recalling $\Pi^r = \sum_{\alpha=1}^N \Pi^r_\alpha  - (v^r)^{-1}$. For equations in the frequency domain, the retarded version has $\omega \to \omega + i \eta$ and $\eta\to 0^+$, and the advanced one is obtained with the Hermitian conjugate.

\section*{appendix C: Derivation of power, torque and force formula in the eigenmode representation}
\setcounter{equation}{0}
\makeatletter
\renewcommand{\theequation}{C\arabic{equation}}
As shown in Fig.~\ref{fig1}(b) in the main text, we assume that the $x$ direction is periodic so that we can transform the stripe in this direction into $k$ space. Then, summing over sites becomes integration over $k$, while keeping the $y$ direction explicitly in real space. We use index $l$ to represent different unit cells in the $x$ direction, $m$ labels the site of carbon atoms in the $y$ direction, and $z$ direction is perpendicular to the graphene plane.

From the main text, the photon's lesser Green's function is defined as
\begin{eqnarray}
	D^<_{\mu\nu}({\bf r'},t';{\bf r},t) = \dfrac{1}{i\hbar}\langle A_\nu({\bf r},t)A_\mu({\bf r'},t')\rangle.
	\label{less}
\end{eqnarray}
Starting from the surface integral, Eq.~(\ref{em}) in the main text [or Eq.~(\ref{eq-main-result}) for the bath at infinity, $\alpha=\infty$] , we can write the emitted energy as
\begin{widetext}
	\begin{equation}
	I = {\rm Re}\frac{1}{\mu_0}\sum_{\mu,\nu,\gamma,\beta,\varsigma}\epsilon_{\mu\nu\gamma}\epsilon_{\gamma\beta\varsigma}\int d\Omega R^2\hat{\bf R}_\mu\int_{0}^{+\infty}\dfrac{d\omega}{\pi}\hbar\omega(-\dfrac{\partial}{\partial x^\prime_{\beta}})D_{\nu\varsigma}^<({\bf r},{\bf r'},\omega)\vert_{{\bf r'}={\bf r}},
\end{equation}
where $\Omega$ is the solid angle. 
Using the Keldysh equation $D^<_{\mu\nu}({\bf r},{\bf r'},\omega)=\sum_{ll',\gamma\varsigma}D^r_{\mu\gamma}({\bf r},{\bf r}_l,\omega)\Pi_{\gamma\varsigma}^{<,ll'}(\omega)D_{\varsigma \nu}^a({\bf r}_{l'},{\bf r'},\omega)$, we have
\begin{equation}
	I = \int_0^\infty d\omega { -\hbar \omega^2 \over 6 \pi^2 \epsilon_0 c^3}{\rm Im}\sum_{ll',\mu}\Pi_{\mu\mu}^{<,ll'}(\omega),
	\label{D1}
\end{equation}
where ${\rm Im}$ takes the imaginary part.
Similarly, we can write the resulting torque of the angular momentum radiation applied to graphene along the $\hat{z}$ direction as
\begin{equation}
	N_z = \int_{0}^{\infty}d\omega\dfrac{\hbar\omega}{6\pi^2\varepsilon_0c^3}{\rm Re}\left[\Pi_{xy}^{<}(\omega)-\Pi_{yx}^{<}(\omega)\right].
	\label{D2}
\end{equation}
Here $\Pi^<$ is the interacting self-energy summed over all lattice sites. For systems with time reversal symmetry, i.e., in which the Hamiltonian is real and symmetric, both torque and force are zero at thermal equilibrium. To generate nonzero torque and force, we have to apply a driven potential bias ($\mu_L$, $\mu_R$) to break the symmetry, and the system is not in equilibrium. Moreover, even in the nonequilibrium state, the nonzero contribution is only from the edge. We apply a periodic boundary condition in the $x$ direction so that the eigenmode can be characterized by traveling waves. The interaction self-energy can be obtained  in the eigenmode as
\begin{equation}
	\Pi_{\mu\nu}^{<,ll'}(\omega)=-i2\pi\sum_{nn'}\langle n\vert M^{\nu l'}\vert n'\rangle\langle n'\vert M^{\mu l}\vert n\rangle f_n(1-f_{n'})\delta(\varepsilon_n-\varepsilon_{n'}-\hbar\omega),
	\label{D3}
\end{equation}
where $n$ denotes modes with wavevectors $k$ and the electron band label and ${\bf M}$ is a vector matrix defined by $e{\bf V}=
\sum_i{\bf M}^i$, which can be expressed explicitly in terms of the matrix elements of velocity, ${\bf M}_{jk}^i=e(\delta_{ij}{\bf V}_{jk}+\delta_{ik}{\bf V}_{jk})/2$, where $i,j,k$ are site indices. 
Substituting Eq.~(\ref{D3}) into Eqs.~(\ref{D1}) and (\ref{D2}) and performing the frequency integration, we get
\begin{equation}
	I = \dfrac{4\alpha}{3\hbar c^2}\sum_{nn'}(\varepsilon_n-\varepsilon_{n'})^2\Theta(\varepsilon_n-\varepsilon_{n'})\sum_\mu\lvert\langle n|V^\mu|{n'}\rangle\rvert^2 f_n(1-f_{n'}),
\end{equation}
and
\begin{equation}
	N_z = \dfrac{4\alpha}{3c^2}\sum_{nn'}(\varepsilon_n-\varepsilon_{n'})\Theta(\varepsilon_n-\varepsilon_{n'}){\rm Im}\left\{f_n(1-f_{n'})\Big[\langle n|V^x|n'\rangle\langle n'|V^y|n\rangle-\langle n|V^y|n'\rangle\langle n'|V^x|n\rangle \Big]\right\}.
\end{equation}
These two equations are the main formulas we used to calculate the energy and angular momentum radiation. On the other hand, the force acting on an object is due to the emission of momentum out of the object. With the Maxwell stress tensor, the force formula integrated over a large sphere surface is equivalent to an integration of the solid angle, i.e., ${\bf F}=\int d\Omega R^2[\varepsilon_0({\hat{\bf R}\cdot E}){\bf E}-u\hat{\bf R}]$. With Eq.~(\ref{less}) and its Fourier transform in space, we obtain
\begin{equation}
	{\bf F} = \int_{0}^{\infty}\dfrac{d\omega}{\pi}\int d\Omega R^2\left[\varepsilon_0(i\hbar\omega^2)\left(D^<\cdot\hat{\bf R}-\frac{1}{2}{\rm Tr}(D^<)\hat{\bf R}\right)+\dfrac{i\hbar}{2\mu_0}{\rm Tr}(\nabla\times D^<\times\stackrel{\leftarrow}{\nabla})\hat{\bf R}\right].
	\label{force}
\end{equation}
Since the only dependence of angle $\Omega$ appears in $\hat{\bf R}$, integration of an odd $\hat{\bf R}$ produces a value of zero. Thus, we must do a dipole expansion of $D^r_{\mu\nu}({\bf R}-{\bf r}_l)$ to have an even order of $\hat{\bf R}$, i.e., $D^r_{\mu\nu}({\bf R}-{\bf r}_l)=D^r_{\mu\nu}({\bf R})-{\bf r}_l\cdot\dfrac{\partial}{\partial{\bf R}}D^r_{\mu\nu}({\bf R})+\cdots$. Substituting it into the Keldysh equation and keeping only the first-order term (ignoring the monopole and higher order terms), we have 
\begin{eqnarray}
	D^<_{\mu\nu} = -\sum_{\zeta,\gamma,\xi,l,l'}\left(D^r_{\mu\zeta}\Pi^{<,ll'}_{\zeta\gamma}x^{l'}_\xi\partial_\xi D^a_{\gamma\nu}+x^l_\zeta\partial_\zeta D^r_{\mu\zeta}\Pi^{<,ll'}_{\zeta\gamma}D^a_{\gamma\nu}\right),
\end{eqnarray}
where all the subscript indices indicate directions $x,y$, and $z$. As the retarded photon Green's function is
\begin{equation}
	D^r_{\mu\nu}\approx-\dfrac{e^{-i\frac{\omega}{c}R}}{4\pi\varepsilon_0c^2R}({\bf U}-\hat{\bf R}\hat{\bf R})_{\mu\nu},
\end{equation}
the Keldysh equation now can be written as 
\begin{equation}
	D^<_{\mu\nu}=-i\frac{\omega}{c}\left(\dfrac{1}{4\pi\varepsilon_0c^2R}\right)^2\sum_{\zeta,\gamma,l,l'}\left[({\bf r}_l-{\bf r}_{l'})\cdot \hat{\bf R}({\bf U}-\hat{\bf R}\hat{\bf R})_{\mu\zeta}\Pi^{<,ll'}_{\zeta\gamma}({\bf U}-\hat{\bf R}\hat{\bf R})_{\gamma\nu}\right].
	\label{keldysh}
\end{equation}
Substituting Eq.~(\ref{keldysh}) into the force formula (\ref{force}) and integrating over the solid angle, we have
\begin{equation}
	F^\mu=\int_{0}^{\infty}d\omega\dfrac{-\hbar\omega^3}{60\varepsilon_0\pi^2c^5}\sum_{ll'}\left\{4{\rm Tr}\left(\Pi^{<,ll'}\right)({\bf r}_l-{\bf r}_{l'})_\mu-\sum_{\nu}\left[\Pi^{<,ll'}_{\mu\nu}({\bf r}_l-{\bf r}_{l'})_\nu+({\bf r}_l-{\bf r}_{l'})_\nu\Pi^{<,ll'}_{\nu\mu}\right]\right\}.
\end{equation}
With Eq.~(\ref{D3}), we have
\begin{equation}
	\sum_{ll'}\Pi^{<,ll'}_{\mu\nu}(\omega)r^l_\gamma=-i2\pi\sum_{nn'}{\rm Tr}\left(\sum_l\langle n|M^{\mu l}r^l_\gamma|n'\rangle\sum_{l'}\langle n'|M^{\nu l'}|n\rangle\right) f_n(1-f_{n'})\delta(\varepsilon_n-\varepsilon_{n'}-\hbar\omega).
\end{equation}
Here we introduce the notation $eU^\mu_\gamma\equiv\sum_l{\bf M}^{l\mu}r^l_\gamma$, where $U^\mu_\gamma$ has two indices of directions $x,y$, and $z$. The superscript index is associated with ${\bf M}^{\mu l}$ or the velocity component, and the subscript index is associated with the direction of the coordinate $r_\gamma$. Then we obtain the final formula for the force,
\begin{eqnarray}
	\nonumber F^\mu &=&\dfrac{4\alpha}{30\hbar^2 c^4}\sum_{nn'}\left(\varepsilon_{n}-\varepsilon_{n'}\right)^3\Theta(\varepsilon_{n}-\varepsilon_{n'})f_n(1-f_{n'}){\rm Tr}\left[4\sum_\nu(\rho_nU^\nu_\mu\rho_{n'}V^\nu-\rho_nV^\nu\rho_{n'}U^\nu_\mu)\right.\\
	&&\left.-\sum_\nu(\rho_nU^\mu_\nu\rho_{n'}V^\nu-\rho_nV^\mu\rho_{n'}U^\nu_\nu)-\sum_\nu(\rho_nU^\nu_\nu\rho_{n'}V^\mu-\rho_nV^\nu\rho_{n'}U^\mu_\nu)\right],
\end{eqnarray}
where $\rho_n=|n\rangle\langle n|$ is the density matrix of state $n$. For a graphene stripe, we find only $F^x\ne 0$, and the other component is identically zero. The group velocity of state $|n\rangle$ is calculated by $\langle n|V^x|n\rangle$, which determines the left or right chemical potential used in the Fermi function $f_n$. 
\end{widetext}

\section*{appendix D: Velocity matrix}
\setcounter{equation}{0}
\makeatletter
\renewcommand{\theequation}{D\arabic{equation}}
Given a Hamiltonian $\hat{H} = C^\dagger HC$ in real space with Hermitian matrix $H_{ij}=H_{ji}^*$ and the creation (annihilation) operator $C^\dagger$ ($C$) for the electrons, the velocity matrix is \cite{zuquan-prb}
\begin{equation}
	{\bf V}_{jk}=\dfrac{1}{i\hbar}H_{jk}({\bf R}_j-{\bf R}_k).
\end{equation}
Let ${\bf v}_1=v(0,1), {\bf v}_2=v(-\sqrt{3}/2,-1/2),{\bf v}_3=v(\sqrt{3}/2,-1/2)$, with $v=a_{cc}t/\hbar$, where $a_{cc} = 0.142$ nm is the bond length between nearest-neighbor carbon atoms in graphene.
For a zigzag graphene nanoribbon, the velocity operator is
\begin{eqnarray}
\nonumber {\bf V}&=&i{\bf v}_1\left[\sum_{l,m}a^\dagger_l(m+1)b_l(m)\right]\\
\nonumber&-&i{\bf v}_2\left[\sum_{l,m=odd}b^\dagger_l(m)a_l(m)+\sum_{l,m=even}b^\dagger_l(m)a_{l-1}(m)\right]\\
\nonumber&+&i{\bf v}_3\left[\sum_{l,m=odd}a^\dagger_l(m)b_{l-1}(m)-\sum_{l,m=even}b^\dagger_l(m)a_l(m)\right]\\
	     &+& h.c.
\end{eqnarray}
Here $m$ represents the site as illustrated in Fig. 1, and $a\ (a^\dagger)$ and $b\ (b^\dagger)$ are annihilation (creation) operators generated by sublattices $A$ and $B$, respectively. Performing a Fourier transform in the $x$ direction with Wakabayashi's convention \cite{Wakabayashi}, the velocity can be written as
\begin{eqnarray}
\nonumber{\bf V}&=&i{\bf v}_1\sum_{k_x,m}a^\dagger_{m+1}(k_x)b_m(k_x)\\
\nonumber&-&i{\bf v}_2\sum_{k_x,m}\xi b^\dagger_m(k_x)a_m(k_x)\\
	&+&i{\bf v}_3\sum_{k_x,m}\xi a^\dagger_m(k_x)b_m(k_x)+{\rm H.c.},
\end{eqnarray}
where $\xi=e^{-ik_x\tilde{a}/2}$ and $\tilde{a}=\sqrt{3}a_{cc}$ is the lattice constant.
Then the velocity matrix can be written as
\begin{equation}
{ V} = \left[
	\begin{array}{cc}
	0&{ u}\\
	{ u}^\dagger&0\\
	\end{array}
\right].
\end{equation}
The explicit expressions for $V^x$, and $V^y$ can be expressed by ${ u}_x$ and ${ u}_y$, which are given by
\begin{equation}
	u^x=\sqrt{3}v\sin\dfrac{k_x\tilde{a}}{2}\left[
	\begin{array}{ccccc}
		1&0&0&\cdots&0\\
		0&1&0&\cdots&0\\
		0&0&1&\cdots&0\\
		\vdots&\vdots&\vdots&\ddots&\vdots\\
		0&0&0&\cdots&1\\
	\end{array}
	\right],
\end{equation}
and
\begin{equation}
	u^y=iv\left[
	\begin{array}{ccccc}
		-g_k/2&0&0&\cdots&0\\
		1&-g_k/2&0&\cdots&0\\
		0&1&-g_k/2&\cdots&0\\
		\vdots&\vdots&\vdots&\ddots&\vdots\\
		0&0&0&\cdots&-g_k/2\\
	\end{array}
	\right],
\end{equation}
where $g_k=2\cos(k_x\tilde{a}/2)$. Similarly, the velocity matrices for armchair graphene are
\begin{equation}
	u^x=iv\left[
	\begin{array}{ccccc}
		\xi&-1/2&0&\cdots&0\\
		-1/2&\xi&-1/2&\cdots&0\\
		0&-1/2&\xi&\cdots&0\\
		\vdots&\vdots&\vdots&\ddots&\vdots\\
		0&0&0&\cdots&\xi\\
	\end{array}
	\right],
\end{equation}
and
\begin{equation}
	u^y=iv\dfrac{\sqrt{3}}{2}\left[
	\begin{array}{ccccc}
		0&-1&0&\cdots&0\\
		1&0&-1&\cdots&0\\
		0&1&0&\cdots&0\\
		\vdots&\vdots&\vdots&\ddots&\vdots\\
		0&0&0&\cdots&0\\
	\end{array}
	\right].
\end{equation}
For a stripe with two edges, the angular momenta cancel as the contributions from each edge are equal in magnitude and opposite in sign. To obtain the effect of one edge, in calculation we use the method of sharp cutoff. We separate the graphene nanoribbon into bottom and top parts. The wave functions of atoms in the top part are set to zero. This is analytically equivalent to setting elements of velocity matrices to zero for atoms in the top part.

\bibliographystyle{apsrev4-2}
\bibliography{PFN-emission.bib}
\end{document}